\documentclass[useAMS,usenatbib]{mn2e}

\usepackage{graphicx}
\usepackage{epsfig} 
\usepackage{multirow} 
\usepackage{amssymb}
\usepackage{aas_macros}

 \title[Intense star formation in the Brightest Cluster Galaxy of the massive cluster MACS1931]
       {Starbursting Brightest Cluster Galaxy: a \textit{Herschel} view of the massive cluster MACS J1931.8--2634}       
\author[Joana S. Santos et al.]
{J. S. Santos$^{1}$\thanks{E-mail:jsantos@arcetri.astro.it}, I. Balestra$^{2}$, P. Tozzi$^{1}$, B. Altieri$^{3}$, I. Valtchanov$^{3}$, A. Mercurio$^{4}$,\and
 M. Nonino$^{2}$, Heng Yu$^{5,6,7}$, P. Rosati$^{8}$, C. Grillo$^{9}$, E. Medezinski$^{10,11}$, A. Biviano$^{2}$ \\
$^{1}$ INAF - Osservatorio Astrofisico di Arcetri, Largo Enrico Fermi 5, 50125 - Firenze, Italy\\ 
$^{2}$ INAF - Osservatorio Astronomico di Trieste, Via Tiepolo 11, 34131 - Trieste, Italy\\ 
$^{3}$ European Space Astronomy Centre (ESAC)/ESA, Villanueva de la Ca\~nada, 28691, Madrid, Spain\\ 
$^{4}$ INAF - Osservatorio Astronomico di Capodimonte, Salita Moiariello 16, 80131 - Napoli, Italy\\ 
$^{5}$ Department of Astronomy, Beijing Normal University, Beijing, 100875 China \\
$^{6}$  Dipartimento di Fisica, Universit\'a di Torino, Via P. Giuria 1, I-10125 Torino, Italy  \\
$^{7}$  Istituto Nazionale di Fisica Nucleare (INFN), Sezione di Torino, Via P. Giuria 1, 10125 Torino, Italy \\
$^{8}$ Department of Physics and Earth Science, University of Ferrara, Via Saragat, 1, 44122 Ferrara, Italy \\
$^{9}$ Dark Cosmology Centre, Niels Bohr Institute, University of Copenhagen, Juliane Maries Vej 30,DK-2100 Copenhagen, Denmark \\
$^{10}$ Center for Astrophysics and Planetary Science, Racah Institute of Physics, The Hebrew University, Jerusalem 91904, Israel \\
$^{11}$ Department of Physics and Astronomy, The Johns Hopkins University,  3400  North  Charles  Street,  Baltimore,  MD  21218, USA}
\date{Accepted ....
       Received ...;  
       }
 \pubyear{2015}

\begin{document}
  \maketitle

  \begin{abstract}
  We investigate the dust-obscured star formation properties of the
 massive, X-ray selected galaxy cluster MACS J1931.8--2634 at
 $z$=0.352.  Using far-infrared (FIR) imaging in the range
 100-500$\mu$m obtained with the \textit{Herschel} telescope, we
 extract 31 sources (2$\sigma$) within $r\sim$1 Mpc from the brightest
 cluster galaxy (BCG). Among these sources we identify six cluster
 members for which we perform an analysis of their spectral energy
 distributions (SEDs).  We measure total infrared luminosity
 (L$_{IR}$), star formation rate (SFR) and dust temperature.  
 The BCG, with L$_{IR}$=1.4$\times$10$^{12}$L$_\odot$
 is an Ultra Luminous Infrared Galaxy and hosts a type II AGN.
 We decompose its FIR SED into AGN and starburst components and find
 equal contributions from AGN and starburst.  We also recompute the
 SFR of the BCG finding SFR=150$\pm$15 M$_\odot$yr$^{-1}$.  We search
 for an isobaric cooling flow in the cool core using {\sl Chandra} X-ray data, and
 find no evidence for gas colder than 1.8 keV in the inner 30 kpc, for
 an upper limit to the istantaneous mass-deposition rate of 58
 M$_\odot$yr$^{-1}$ at 95\% c.l.  This value is $3\times$ lower than
 the SFR in the BCG, suggesting that the on-going SF episode lasts longer than the ICM cooling events.
  \end{abstract}


  \section{Introduction}

The cores of galaxy clusters are ubiquitously populated by old, passively evolving spheroids, with little evidence for ongoing or recent episodes of star 
formation \citep[e.g.,] [] {Dressler80, Dressler97, vonderlinden, Girardi}.
This suppression of star formation (SF) is mainly caused by interactions among the densely packed galaxies 
\citep[e.g.,] [] {Moore, Gnedin}, and to a lesser extent by 
interactions between the hot, X-ray emitting intracluster medium (ICM) and the galaxies. 
The brightest cluster galaxy (BCG) that usually sits at the bottom of the potential well and is coincident with the peak of the cluster X-ray emission, 
is typically a very massive, bright, early type galaxy, that only rarely is associated with significant star formation activity \citep[e.g.,] [] {Samuele, Rawle, Fogarty}.

Cool core (CC) clusters are systems whose ICM shows a 
minimum core temperature that is about one third of the global ICM temperature and a low 
core entropy ($<$30 keV~cm$^{2}$), that reflects significant radiative cooling taking place in the cluster 
innermost regions \citep[e.g.,][]{Peterson, Hudson}.  Observations have shown that BCGs with ongoing star
 formation activity are usually hosted by CC clusters \citep{Hoffer}.  However, there is still a large variance in 
current results on the fraction of star forming BCGs and the amount of their star formation rate (SFR). 
This is partly because different diagnostics are used (e.g., optical emission lines, UV continuum, far-infrared)
that may be affected by dust emission and AGN contamination, but also because samples are often not
representative.  In particular, \cite{Samuele} investigated the star formation activity in a sample of 77 BCGs
drawn from a flux limited, X-ray selected cluster sample and reported a lack of star formation in that sample,
based only on optical emission lines.  In contrast, \cite{Rawle} detected star formation in 15 out of 68 BCGs
using a more robust diagnostic based on the FIR emission. The caveat in this fraction is that the sample of 
BCGs originate from a mix of cluster samples, mostly selected to include massive and relaxed clusters 
that are thus biased toward CCs, therefore are more likely associated with star forming BCGs.

It is well known that AGN play a crucial role in the regulation of the star formation in BCGs 
\citep[e.g.,][]{Hlavacek-Larrondo, Russell}.  Ample evidence for AGN feedback has been collected in the last
 decade, where radio jets have been shown to inflate bubbles in the ICM and 
hence offset cooling, nonetheless the link between AGN and star formation is not yet properly established 
\citep[e.g.,] [for a review] {McNamara}.  The recent study of the Phoenix cluster at $z$=0.596 
\citep{McDonald}, the strongest CC cluster known to date, showed a BCG with a very high SFR \textit{and} an
 equally significant AGN activity. 

The X-ray selected cluster MACS J1931.8--2634 \citep [MACS1931 hereafter,][]{Ebeling}
at $z$=0.352 is part of the CLASH  \citep[Cluster Lensing And Supernova survey with Hubble,][]{Postman}
sample of 25 massive clusters used to study the distribution of dark matter in clusters. 
As all of the CLASH clusters, MACS1931 is massive 
 with M$_{200}$=9.9$\pm$0.7$\times$10$^{14}$ M$_\odot$ \citep{Merten}
with a relaxed X-ray morphology,
and harbors a cool core \citep{Ehlert}. The cluster is dominated by a very large, luminous central galaxy 
that, contrary to what is common in most massive clusters, is undergoing a phase of copious star formation. Measured star formation rates range from 
80 M$_\odot$yr$^{-1}$ \citep[rest-frame UV imaging, ] [] {Donahue} to 170 M$_\odot$yr$^{-1}$ \citep[broad band optical imaging,] [] {Ehlert}.
However, these SFRs may be contaminated by AGN activity and underestimated due to dust obscuration. 

The work presented in this Letter aims to overcome these two biases. We present the analysis of \textit{Herschel} \citep{Pilbratt} 100--500 $\mu$m
 observations of MACS1931 that cover the peak of the SED of starbursts.   
 The far-infrared (FIR) is the best diagnostic for star formation as it provides a direct measure of the reprocessed UV light from the on-going star formation, allowing us to measure the total FIR 
 luminosity, star formation rates and dust temperatures of the cluster members. We also 
focus on  the BCG and its environment using {\sl Chandra} X-ray data.
The work presented here is part of a larger \textit{Herschel} study including all CLASH clusters. 
The cosmological parameters used throughout the paper are: $H_{0}$=70 $h$ km/s/Mpc,
$\Omega_{\Lambda}$=0.7 and $\Omega_{\rm m}$=0.3. 

  \begin{figure}
\includegraphics[width=8.2cm,angle=0]{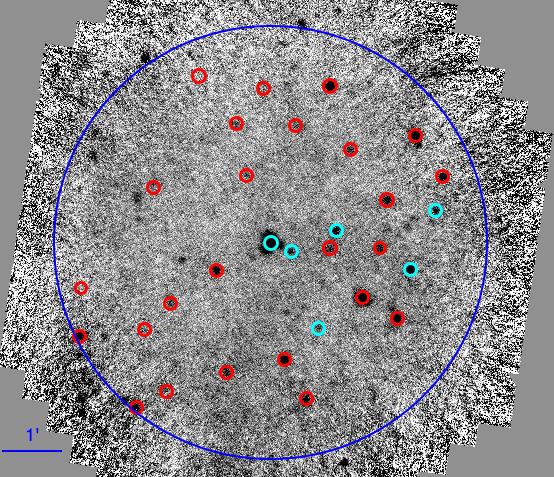} 
\vspace{-0.2cm}
 \caption{PACS 100$\mu m$ image of the cluster field centered on the BCG. The blue circle with 
 $r \sim$1 Mpc ($\sim$1/2 of $r_{200}$)
  indicates the region of source extraction. The small (6$\arcsec$ radius) circles represent the 31 
 individual FIR sources in our catalog (cluster members are highlighted in cyan). North is up and East to the left.}
 \label{herschelimage}
 \end{figure}

  \section{Data}

Although the present work is focused on data from the \textit{Herschel} space telescope, we use ample ancillary data both proprietary and archival: 
mid-infrared data from WISE; X-ray data from \textit{Chandra}; optical data from Subaru (BVR$_{c}$I$_{c}$z), the 
\textit{Hubble Space Telescope} (HST), and extensive VLT/VIMOS spectroscopy. 
  
  \subsection{\textit{Herschel} observations and data reduction}
 
The \textit{Herschel} observations of MACS1931 were carried out in 2011, 2012 and 2013 as 
Open Time 1 and 2 programmes 
(PI Egami, obsid = 1342215993, 1342241619, 1342241681, 1342254639) aimed at 
studying the star formation properties of 
lensing clusters. The PACS \citep{Poglitsch} observations at 100 and 160$\mu$m were performed 
in scan map mode.
The maps were produced using Unimap \citep{Piazzo}: a Generalized Least Square map-maker, 
that allows us to reach ultimate sensitivity with no flux loss, and without iterative masking of the sources.
The 1$\sigma$ noise of the maps is 1.6 mJy in the 100 $\mu$m band and 3.5 mJy in the 160  $\mu$m image.
SPIRE \citep{Griffin} maps with a $\sim  5 \arcmin$ radius were obtained in small map mode.
The SPIRE maps at 250, 350 and 500 $\mu$m with nominal pixel sizes of 6$\arcsec$, 10$\arcsec$ and 
14$\arcsec$, respectively, are dominated by  confusion noise with an \textrm{rms} in the center of 6.2, 6.5 and 7.3 mJy.

\begin{table*}
\caption{Properties of the FIR cluster members, with spectroscopic redshift and photometric redshift 
concordant with the cluster.}  
\label{table:1}     
\small
\centering           
\begin{tabular}{llllllllll}
\hline\hline   
ID         &  RA    &        DEC  &         $z$       &   r$_{proj}$              &    F$_{100\mu m}$  &  F$_{160\mu m}$   &   LIR                                      &  SFR                        & T$_{dust}$   \\
             &           &                  &                     &        kpc                   &     mJy                     &     mJy                    &     $\times 10^{11} L_\odot$  & M$_\odot$yr$^{-1}$  & K   \\
\hline                            
     62       &  292.905501 � &�-26.5669131         &   0.3644    &  923   &   6.5$\pm$1.7         &    9.6$\pm$3.6              &       0.43$\pm$0.04     &     6.4$\pm$0.6    & 29$\pm$5     \\
     69 (BCG) &  292.956707   &  -26.5758907   &  0.352     &    --    &   212.5$\pm$15.0   &    231.3$\pm$16.5   &       14$\pm$2             &     210$\pm$23   &  33$\pm$2     \\
     75*       &  292.950604 ��&�-26.5782649         &  0.3652     & 117   &   3.2$\pm$1.6         &    10.7$\pm$3.6            &       0.47$\pm$0.07     &    7.0$\pm$1.5    & 14$\pm$1   \\
     89        &  292.941855   &  -26.5994976       & 0.3494      &  498     &   4.0$\pm$1.6         &    10.1$\pm$3.6        &       0.34$\pm$0.04     &    5.0$\pm$0.5    & 24$\pm$4      \\
  \hline
   68         &  292.936482 � &� -26.5724383  &  0.36$\pm$0.07   &     366      &   8.9$\pm$1.7          &    12.4$\pm$3.6     &       0.54$\pm$0.06     &    8.0$\pm$1.0    & 30$\pm$4     \\ 
   80         &  292.913344  �&� -26.5832069  &  0.34$\pm$0.07    &     785      &    21.5$\pm$2.2   &    32.2$\pm$4.2    &       1.3$\pm$0.1        &    19.5$\pm$1.7    & 29$\pm$2     \\
  \hline  
\end{tabular}
\vspace{-0.2cm}
   \flushleft  \hspace{0.5cm}   * SPIRE fluxes of ID 75 may be contaminated by the BCG that is located at $\sim$20$\arcsec$ distance.
\end{table*}

\subsection{Far-infrared sources in MACS1931}

Our far-infrared observations of MACS1931 covers a region with 3.6$\arcmin$ radius centered 
on the X-ray cluster center, where the sensitivity of the PACS maps is robust (see Fig. 1). This radius
corresponds to 1.1 Mpc in physical units which is about 1/2 of the cluster virial radius measured from 
lensing \citep[$r_{200}$=1.82$\pm$0.04 Mpc, ][]{Merten}.

Here we outline our procedure to obtain the catalog of FIR sources in the field of MACS1931. 
Our catalog is based on blind source detections in the 100 and 160$\mu m$ maps separately, using SExtractor \citep{Bertin}. 
As standard for PACS data, the photometry is made with fixed apertures with radii of 6$\arcsec$ and 9$\arcsec$ at 100$\mu m$ and 160$\mu m$, respectively, 
corrected with the encircled energy factors given by \cite{Balog}. This procedure was validated with manual aperture photometry.
Given the difficulty to obtain reliable errors with standard source detection algorithms because of the correlated noise present in PACS data, 
we compute the photometric errors as the 1$\sigma$ detection limits in each band, in addition to 7\%
(calibration accuracy of the flux scale) of the source flux. 
The SPIRE source detection was performed using a simultaneous fit to all sources in the prior list 
based on the PACS detections. We run the XID method \citep[][]{Roseboom} 
using the same prior catalog on the three SPIRE bands, using the 
corresponding SPIRE point response function for each band.
If the fitted SPIRE flux density at the position of an input PACS source is below the $3\sigma$ sensitivity in each band we assigned the $3\sigma$ 
values as upper limits. These correspond to 3$\times$ the confusion noise and are  equal to 17.4, 18.9 and 20.4 mJy at 250, 350 and 500$\mu m$, 
respectively.

We obtain 31 detections at $>$2$\sigma$ in at least one of the PACS bands, within $r$=3.6$\arcmin$ 
centered on the BCG. To identify the origin of these sources we match the FIR catalog with our spectroscopic and photometric
redshift catalogues. The spectroscopic catalog consists of 2800 redshifts obtained with VIMOS \citep[CLASH-VLT Large Programme 186.A-0798, PI Rosati,][]{Rosati}, 
whereas the photo-$z$ catalogues are based on photometry from the HST and Subaru\footnote{CLASH catalogs: https://archive.stsci.edu/prepds/clash/}.
We find that, of the 31 sources in the cluster field, 4 are confirmed members, 2 are candidate members (photo-$z$) and 
18 are interlopers. Since the completeness of our spectroscopic sample of cluster members is close to 90\% it is unlikely that the remaining 7 
\textit{Herschel} sources are at the cluster redshift.

  \section{Far-infrared properties of the cluster galaxies}

We fit the galaxies FIR SEDs using {\tt LePhare}  \citep{Arnouts} with Chary \& Elbaz (2001) templates, to
 measure the galaxy integrated infrared luminosity $L_{IR}$ in the range 8--1000$\mu$m. 
The star formation rates are derived using the updated scaling relation, 
SFR$_{IR}$ = 1.48$\times$10$^{-10} L_{IR}/ L_\odot$, 
\citep{Kennicutt2012, Murphy, Hao} that uses a Kroupa initial mass function \citep{Kroupa}\footnote{These SFRs are lower than the ones 
obtained with the widely used Kennicutt 1998 calibration (SFR=0.86$\times$SFR$_{K98}$)}.
The SFR of our sample, measured with pure starburst templates, spans the range 
5 - 210 M$_\odot$yr$^{-1}$.  If we exclude the 
highest star-forming galaxy - the BCG - we find an average SFR of 9.2 M$_\odot$yr$^{-1}$. 
The total SFR of the six cluster galaxies amounts to 256 M$_\odot$yr$^{-1}$.  
This value is in  good agreement with the recent result on the SFR of massive clusters using 
\textit{Herschel} data by \cite{Popesso}.
In the next section we perform a more detailed study of the BCG and refine its SFR measure, after 
accounting for the impact of the AGN.
With the exception of the BCG that is an Ultra Luminous Infrared Galaxy 
(ULIRG, $\gtrsim$10$^{12} L_\odot$), the FIR detected cluster galaxies are 
LIRGs or normal star forming galaxies. 

The temperature of the dust, $T_{dust}$, present in the galaxies is computed with a modified black body 
model with an emissivity index $\beta$ 
fixed to 1.5. Apart from galaxy ID 75, that shows some contamination with the BCG fluxes in the 
SPIRE bands, we find $T_{dust}$ 
in the range 24-33 K, which is within the range of dust temperatures for $z\le$0.3 LIRGs and 
ULIRGs \citep{Magdis}.  The FIR properties of the cluster members are summarized in Table 1.
  
\section{Connection between the BCG, the AGN and the ICM}

In this section we analyze in detail the BCG sitting at the core of the cluster. It is uncommon to have a 
cluster with a massive BCG with a SFR level of 210 M$_\odot$yr$^{-1}$ \citep{Rawle, Hoffer}. 
However, our initial SFR result may be biased by the presence of a strong obscured AGN, which needs 
to be carefully modeled.  We also explore the interconnections between the star formation of 
the BCG and the mass deposition rate in the cluster core.
The UV and optical properties of this galaxy have been studied in \cite{Donahue} and \cite{Ehlert}, 
that computed the galaxy star formation rate using different diagnostics. We convert their values to 
the ones obtained with the updated calibration used here, as described in 
\cite{Kennicutt2012}.  While \cite{Donahue} found a SFR = 69 M$_\odot$yr$^{-1}$ using UV 
photometry, \cite{Ehlert}, in a crude approximation,  derived a value twice larger, 
146 M$_\odot$yr$^{-1}$, based on optical broad band imaging. 
These calculations do not account for dust extinction nor for contamination from AGN activity.  

\subsection{SED decomposition: AGN \& star formation}
 
\begin{figure}
\includegraphics[width=7.5cm,angle=0]{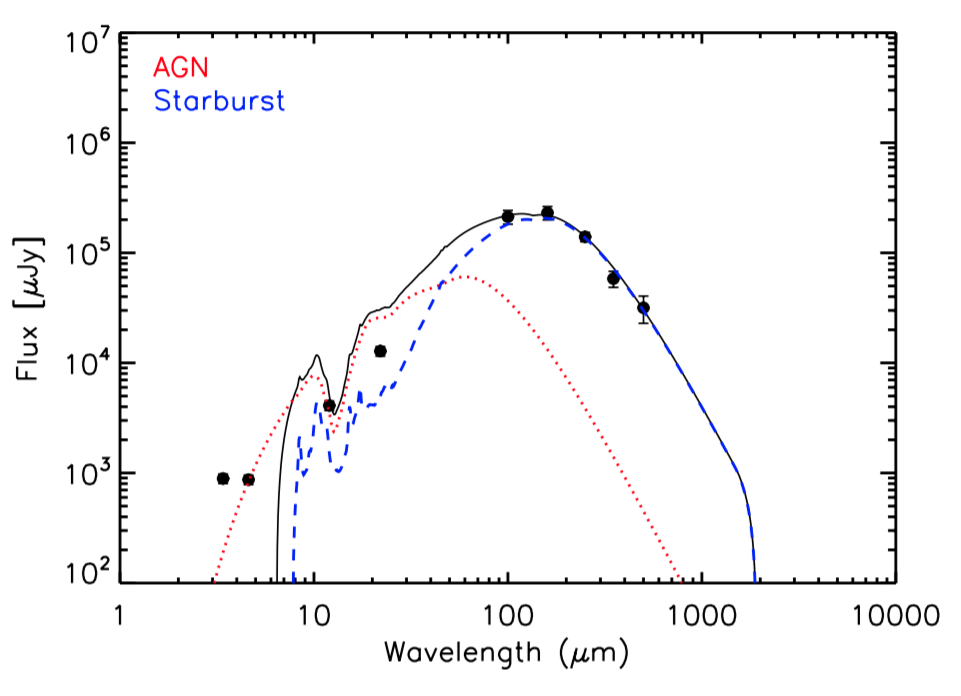} 
\vspace{-0.4cm}
\caption{Infrared SED of the BCG. Black points show the observed data from WISE and \textit{Herschel} 
and the black solid line represents the best-fit SED model. 
AGN and starburst components from the combined SED are shown in red and blue, respectively.}
\label{bcg_sed}
\end{figure}

The BCG of MACS1931 hosts an X-ray bright AGN, embedded in the ICM emission, which we model and 
subtract from the AGN signal.  We detect about 1030 counts in the 0.5-7 keV band. 
Our spectral analysis with an intrinsic slope fixed to $\Gamma = 1.8$, provides an intrinsic absorption of
$N_{H}=1.94_{-0.19}^{+0.21} \times 10^{22}$ cm$^{-2}$, which is very close to the canonical 
value $10^{22}$ cm$^{-2}$ above which an AGN is classified as absorbed. 
The  unabsorbed rest-frame luminosities are $L_{0.5-2 keV} = 3.6 \times 10^{43}$ ergs s$^{-1}$  
and  $L_{2-10 keV} = 6.9 \times 10^{43}$ ergs s$^{-1}$.
 
We first investigate the contribution of the AGN component to the FIR emission using 
{\tt DecompIR} \citep{Mullaney}, 
an SED model fitting software that decomposes the FIR SED in AGN and starburst components.
In short, the AGN component is an empirical model based on observations of moderate-luminosity local AGNs,
whereas the 5 starburst models represent a typical range of SED types, with an extrapolation beyond 
100$\mu m$ using a grey body with $\beta$=1.5. 
The best-fit model obtained with {\tt DecompIR} considering the \textit{Herschel} datapoints (Fig. 2) yields
$L_{IR}$ = 2.2$\times10^{12}$ L$\odot$, with AGN and starburst contributions of 53\% and 47\%, respectively.
This allows us to recompute the SFR removing the AGN contamination.  We thus obtain SFR(BCG)= 
150$\pm$15 M$_\odot$yr$^{-1}$, a value similar to that reported in \cite{Ehlert} but much more robust.

\subsection{Mass deposition rate and SFR}

MACS1931 harbors one of the most X-ray luminous cool cores known.  
The properties of the core of MACJ1931 have been investigated in detail in \citet{Ehlert}, where an equivalent
mass deposition rate of $\sim 700 M_\odot$ yr$^{-1}$  in the inner 70 kpc has been estimated
in the assumption of isobaric cooling.  Despite the  high cooling rate, this cluster is missing the central
metallicity peak which is otherwise measured in the majority of CC clusters  
\citep{DeGrandi}. This suggests bulk transport of cool gas out to large distances from the centre 
due to the powerful AGN outburst \citep{Ehlert}.  In particular, a bright, dense region north of the 
BCG shows  low-temperature and high-density metal-rich gas and is consistent with being a 
remnant of the cool core after it was disrupted by the AGN \citep[see also][]{Kirkpatrick}.

\begin{figure*}
\vspace{-0.2cm}
\hspace{-0.5cm}\includegraphics[width=7.7cm]{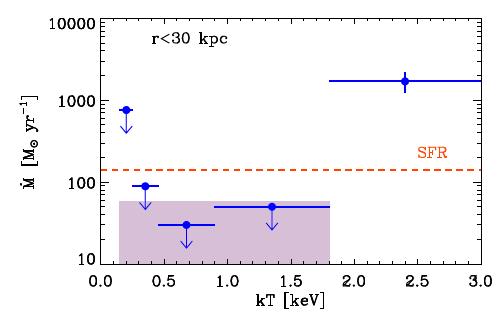}
\hspace{1.0cm}\includegraphics[width=5.cm]{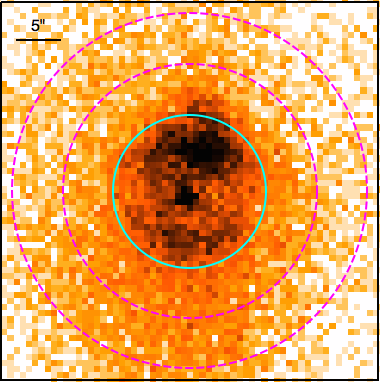}
\caption{ \textit{Left} $\dot M / M_{\odot}$ (blue circles) measured in different temperature bins in the 
inner 30 kpc of MACJ1931. Errors bars correspond to 1$\sigma$ while arrows indicate single-sided 
95\% upper limits. The shaded area is the single-sided 95\% upper limit on the global mass cooling 
rate obtained with a single {\tt mkcflow} model in the temperature range 0.15 --1.8 keV. 
The horizontal line marks the star formation rate of the BCG.  \textit{Right} Soft X-ray band 
(0.5$\arcmin \times$0.5$\arcmin$) centered on the BCG.   The regions in which we computed $\dot M$ 
are shown in cyan (inner 30 kpc) and in magenta (annulus with 50 -- 70 kpc radius).
North is up and East to the left.}
\end{figure*}

We constrain the ICM mass cooling rate, $\dot M$, in the inner 30 kpc 
around the BCG under the same assumption of isobaric cooling.
The main difference relative to the previous analysis by \cite{Ehlert} is that we focus on a 
limited temperature range, 0.15-3.0 keV, and in particular below 1.8 keV.  In fact, 
the signature of an isobaric cooling flow is given by a specific relation between the emission measure 
and the gas temperature, as described in the model {\tt mkcflow} \citep{Mushotzky}.
Therefore we measure the cooling rate of the gas independently
in five temperature bins, using a set of {\tt mkcflow} models 
within XSPEC \citep{Arnaud} in the following temperature intervals: 0.15--0.25, 0.25--0.45, 0.45--0.9,
 0.9--1.8,1.8--3.0 keV.  We also consider a single temperature 
 {\tt mekal} \citep{Mewe85,Mewe86,Kaastra,Liedahl} component to account for the gas 
hotter than 3 keV.  We detect about 13700 net counts (0.5-7.0 keV band) in the inner 30 kpc.  
Since it is not possible to measure the metal abundance of the cold gas, given 
its low emission measure, we conservatively assume that 
its metallicity is equal to that of the {\tt mekal} component which dominates the emission,
which is $Z = 0.35 \pm 0.05 Z_\odot$.  We find that the mass deposition rate for gas below 3 keV has a
95\% single-sided upper limit of $\dot  M < 135M_\odot$ yr$^{1}$ (see dashed horizontal line 
in Fig. 3, left panel),
while the temperature of the {\tt mekal} component is $kT = 6.43^{+0.50}_{-0.45}$ keV.
When we try to constrain $\dot M$ in temperature intervals, we find that we are able to measure
a substantial mass deposition rate only in the temperature range 1.8-3.0 keV.
Below 1.8 keV, the upper limit on the mass deposition rate is only $\dot M < 58 M_\odot$ yr$^{1}$ 
at a 95\% c.l.  The upper limits measured in different temperature bins are shown in Fig. 3, left panel. 
We find that $\dot M$ is significantly lower - at least by a factor 3 - than the measured SFR. 

This result is in broad agreement with the correlation between the SFR observed in the BCG and the 
properties of the hosting cool core \citep{Rafferty, Rawle}, but in disagreement with previous measurements 
of a mass deposition rate typically higher than the SFR in the BCG
\citep{Odea,McDonald2011,Mittal}.  Clearly, the ratio of the mass deposition rate and the SFR is
sensitive to the times scales for  ICM cooling and star formation and it is not expected to vary much
among clusters.  To check whether we can reconcile the results for MACSJ1931 with previous results, 
we also explore the presence of colder clouds (with temperature $< 10^7$ K) in the ICM
beyond 30 kpc, where a significant fraction of colder gas can cool and fall into 
the innermost regions.  This is suggested by a recent work by \cite{Voit} which showed that colder clouds 
can precipitate out of the hot gas via thermal instability on a large region centered in the core, 
feeding black hole accretion and/or star formation in the BCG.  Even in this case we find an upper limit 
(95\% c.l.) of 45 and 52 $M_\odot$  yr$^{-1}$ in annuli of 30-50 and 50-70 kpc, respectively.
Therefore, we do not find evidence of a large amount of colder gas within 30 kpc nor 
falling from beyond 30 kpc onto the centre.
The low upper limits on the mass deposition rate found in the case of MACSJ1931, may require a 
scenario where the time scale for star formation rate in the BCG is longer than the 
cooling events occurring intermittently in the
cluster core.   To investigate this possibility, we are currently exploring the relation between $\dot M$
and SFR in a small sample of nearby clusters whose BCG shows significant star formation 
(Molendi et al. in preparation).



  \section{Conclusions}

In this Letter, we present a study of the dust obscured star formation properties in the galaxy cluster 
MACS1931 at $z$=0.352, and a detailed investigation on the origin of the BCG SFR.
We detect  FIR emission in 6 cluster members and we derive SFRs based on broad band FIR SEDs in the range 
5 - 150 M$_\odot$yr$^{-1}$, and dust temperatures in the range 24-33 K. The strongest \textit{Herschel}
source in the cluster field is, notably, the BCG, one of the few ULIRGs detected in the nearby Universe. 
Our analysis shows the importance of adequately assessing the contamination from AGN, but even 
accounting for $\sim$50\% of L$_{IR}$ due to the AGN we estimate a SFR=150 M$_\odot$yr$^{-1}$ for 
the BCG, a surprising value for a galaxy expected to be "red \& dead".  At similar redshift only Abell 1835 has 
a comparable SFR.   At the same time, cold gas ($kT<$1.8 keV) is not detected in the ICM gas, and the mass 
deposition rate has a robust upper limit of 58 M$_\odot$yr$^{-1}$, suggesting that in this case star 
formation lasts longer than the ICM cooling events.

 \vspace{-0.2cm}    
     
  \section*{Acknowledgements}
  \small
  We thank G. Risaliti, G. Giovannini, E. Liuzzo and  M. Massardi for useful discussions on the AGN properties.  
We also thank the anonymous referee for useful comments.  
  JSS and IB acknowledge funding from the European Union Seventh Framework Programme (FP7/2007-2013) 
  under grant agreement nr 267251 "Astronomy Fellowships in Italy" (AstroFIt).
We acknowledge financial support from PRIN INAF 2012 {\it A unique dataset
to address the most compelling open questions about X-ray galaxy clusters} and 
PRIN-INAF 2014: {\it Glittering Kaleidoscopes in the sky, the multifaceted nature and role of galaxy clusters}. 
Results are partly based on ESO LP186.A-0798. 
Herschel is an ESA space observatory with science instruments provided by European-led PI 
consortia and with important participation from NASA.

\vspace{-0.2cm}


\end{document}